\documentclass{mn2e}
\usepackage[utf8x]{inputenc}
\usepackage{graphicx}
\usepackage{natbib}
\usepackage{amssymb}
\usepackage[usenames,dvipsnames]{xcolor}
\usepackage{astrojournals}
\newcommand{\Msun}{M_\odot}
\newcommand{\lcdm}{$\Lambda$CDM}

\title{Can Feedback Solve the Too Big to Fail Problem?}
\author[S. Garrison-Kimmel et al.]{Shea Garrison-Kimmel\thanks{$\!$sgarriso@uci.edu},
  Miguel Rocha,
  Michael Boylan-Kolchin\thanks{$\!$Center for Galaxy Evolution
  fellow},\newauthor
  James S. Bullock,
  Jaspreet Lally\\
  \noindent$\!\!$Center for Cosmology, Department of Physics and Astronomy,
  University of California, Irvine, CA 92697, USA}
\begin{document}

 \pagerange{\pageref{firstpage}--\pageref{lastpage}} 
 \pubyear{2013}

\maketitle

\label{firstpage}
\begin{abstract}
  The observed central densities of Milky Way dwarf spheroidal galaxies (dSphs)
  are significantly lower than the densities of the largest ($V_{\rm max} \sim
  35$ km/s) subhalos found in dissipationless simulations of Galaxy-size dark
  matter hosts. One possible explanation is that gas removal from feedback can
  lower core densities enough to match observations.  We model the dynamical
  effects of supernova feedback through the use of a time-varying central
  potential in high resolution, idealized numerical simulations and explore the
  resulting impact on the mass distributions of dwarf dark matter halos.
  We find that in order to match the observed central masses of $M_\star \sim
  10^6 \Msun$ dSphs, the energy equivalent of more than $40$,$000$ supernovae
  must be delivered with 100\% efficiency directly to the dark matter.  This
  energy requirement exceeds the number of supernovae that have ever exploded in
  most dSphs for typical initial mass functions.  We also find that, per unit
  energy delivered and per cumulative mass removed from the galaxy, single
  blow-out events are more effective than repeated small bursts in reducing
  central dark matter densities. We conclude that it is unlikely that supernova
  feedback alone can solve the ``Too Big to Fail'' problem for Milky Way
  subhalos.
\end{abstract}


\section{Introduction}
\label{sec:intro}
The current paradigm for structure formation, cold dark matter with a
cosmological constant ($\Lambda$CDM), has proven successful at reproducing the large
scale universe \citep[][and references therein]{Hinshawetal2012,Hoetal2012};
however, disparities exist between the theory and observations on small
scales. For example, the rotation curves of dwarf and low surface brightness (LSB)
galaxies appear to favor core-like density distributions rather than the
cuspy distributions seen in dissipationless simulations \citep{Flores1994,
  Moore1994, kuzio-de-naray2008, Trachternachetal2008, deBlok2010, kuzio2011}.
There has been much discussion in the literature regarding the ability of
baryonic processes, i.e. feedback, to displace dark matter and resolve the
problem \citep{Navarro1996, Mashchenko2006, Pontzen11, Ogiya2012,
  teyssieretal2012} -- such arguments seem reasonable given the fairly large stellar mass
($M_\star \sim 10^8M_\odot$) of a typical LSB galaxy.

Perhaps more troubling is that a similar problem appears to exist for even lower
luminosity dwarf spheroidal (dSph) galaxies ($M_\star \sim 10^6 M_\odot$) in the Local
Group. \citet{Walker2011}, \citet{Jardel2012}, \citet{Agnello2012}, and 
\citet{Amorisco2013}, among others, find
evidence for cores in the Fornax and Sculptor dSphs.  This is particularly important if
true, as the same mechanisms shown to
to flatten dark matter profiles in larger galaxies appear to have
little effect in galaxies with so few stars  \citep{Governato2012}.
The density profiles of
these dSphs is a matter of active debate, however:  \citet{Breddels2013}
argue that it is unlikely that Fornax, Sculptor, Carina, and
Sextans are embedded in cored dark matter profiles, and Strigari,
Frenk, and White (in preparation) find that it is indeed possible to match the
kinematics and photometry of Fornax and Sculptor in cuspy dark matter
potentials for generalized forms of the stellar density distribution and stellar velocity anisotropy profile.

Independent of the functional shape of the dark matter density profile of the
Milky Way dSphs, it has become clear that the dark matter masses of the dSphs
are significantly lower than expected for the most massive subhalos in
dissipationless \lcdm\ simulations (the so-called ``too big to fail'' (TBTF)
problem;
\citealt{Boylan-Kolch11,Boylan-Kolch2012}).  
The core/cusp and Too Big To Fail problems may be closely related: if the
majority of the bright dSphs in the Milky Way have dark matter cores of
$500-1000$ pc, then their central masses would be reduced by a factor of 2-3,
precisely the amount that is required to explain TBTF. If most or all of the
dSphs have non-cored profiles, however, the two issues are distinct.

\begin{figure}
  \centering
    \includegraphics[width=0.45\textwidth]{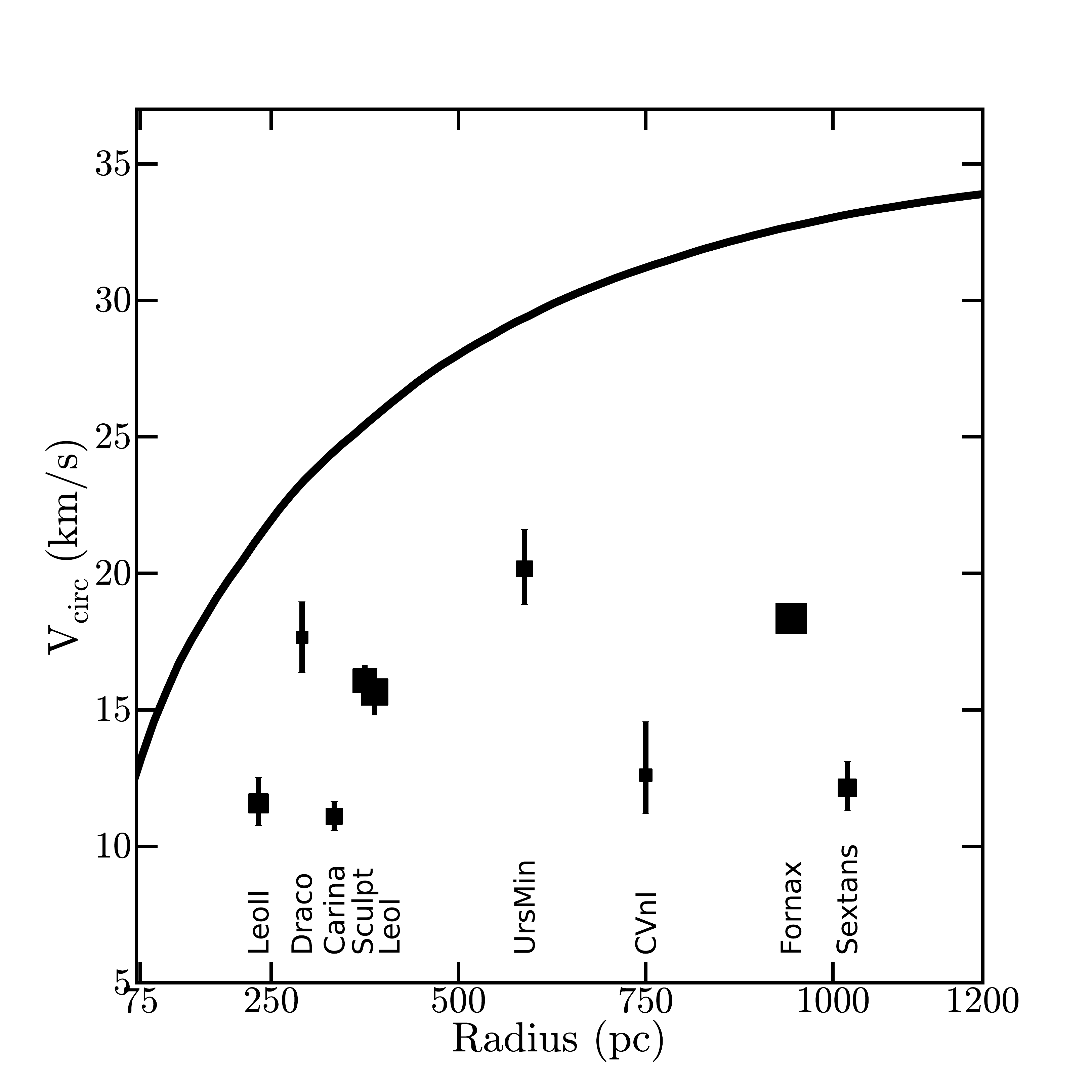}
    \caption{The simulated circular velocity profile as a function of subhalo
      radius along with observed circular velocities.  The solid line shows the
      circular velocity profile of our idealized halo in the initial conditions,
      which is representative of the largest subhalos found in dark-matter-only
      simulations of Milky Way-size halos, a Too Big to Fail subhalo. The
      observational values are data for the bright dSphs ($L_V >
      10^{5}\,L_{\odot}$; see text for details) and the size of each point is
      proportional to the luminosity of that satellite.  The Milky Way
      satellites have significantly less mass near their center than the halos
      in which abundance matching predicts they form.}
\label{fig:ICs}
\end{figure}

The TBTF problem is illustrated in Figure~\ref{fig:ICs}, in which a circular
velocity profile typical of one of these ``massive failures" is plotted along
with observed values of the mass enclosed within the deprojected half-light
radius of each of the bright dSphs around the Milky Way (computed by
\citealt{Wolf2010}, who used data from \citealt{Walker2009} in addition to
data from \citealt{Munoz2005,Koch2007,Simon2007} and \citealt{Mateo2008}).  If the
largest subhalos do host the bright dwarfs, then the dark matter must be less
dense in their centers than predicted in dissipationless CDM simulations, possibly
because of a combination of star formation feedback, tidal interactions, and
ram-pressure stripping \citep[e.g.][]{Brooks2012,Arraki2012}, or non-standard dark
matter physics \citep[e.g.][]{Lovell2012,Anderhalden2012, vogelsberger12,
Rocha2012}.  Other authors have pointed out that a statistically 
rare or low-mass Milky Way \citep[e.g.][]{Purcell2012, Wang2012} can also explain
the discrepancy; however, the former option is called into question by 
\citet{Strigari2012} and the latter is in conflict with constraints derived 
by \citet{Boylan-Kolchin2012LeoT} from the orbit of Leo I.

While feedback appears a plausible solution to the cusp/core problem in brighter
dwarf galaxies and LSBs --- and more generally, to reducing their central dark
matter content relative to predictions from dissipationless \lcdm\ simulations
--- dSph galaxies are much more dark-matter dominated, with observed
mass-to-light ratios within the stellar radius in excess of 100 in some cases
\citep[e.g.][]{Walker2012}.  Moreover, according to the theoretical
extrapolation of abundance matching, we expect the stellar mass to drop by $\sim
2.5$ dex for a difference of only one decade in halo mass at these mass scales
\citep{Behroozi12}.  The expectation is that the dark matter's gravitational
potential should overwhelm that of the baryons, even at the centers of halos.
Finally, the fact that these systems are so deficient in stars means that the
total energy available to alter the gravitational potential is minimal.

Some groups \citep[e.g.][]{Governato2012,DelPopolo2012} have successfully 
reproduced the low central densities of LSBs by invoking supernovae feedback 
in cosmological zoom-in simulations.  Others 
\citep[e.g.][]{Read2005,Zolotov2012,teyssieretal2012} have managed to
flatten the central density profiles of larger dwarf galaxies 
($M_{\star} \sim 10^7 - 10^8\Msun$) via similar
techniques.   
Reduced central densities are not generic outcomes of simulations 
including gas physics, however: other groups find profiles that are either 
unchanged \citep{Parry2012} or steeper than those in the dark-matter-only 
case owing to adiabatic contraction \citep{diCintio2011}.  Such studies
typically rely upon hydrodynamical sub-grid models, which may be
responsible for these divergent outcomes.  Our approach is complementary to
these fully self-consistent approaches in that we focus on the effect that blowouts have
on centrally located dark matter without regard to the chain of mechanisms responsible for
blowing out the gas.

Compared to studies of LSB galaxies, moreover,  dwarf spheroidals present a much more difficult 
problem numerically owing to their small physical size $\sim 300$~pc.  Indeed, the central
regions of dSph-size subhalos remain extremely difficult to resolve even in
collisionless zoom simulations \citep{Boylan-Kolch2012}, let alone
hydrodynamical simulations: the mass within $\sim 4-5$ force resolution elements
is systematically understimated by $20\%$ because of the gravitational softening
adopted in simulations \citep{Font2011}. 
Poor resolution can give rise to
two-body relaxation errors that tend to flatten the inner density profile, and
this undesired effect propagates radially outwards in the cumulative velocity
profile. Moreover, if the dark matter potential is shallower due to the lack of
adequate resolution, gas outflows and tidal effects may over-predict the removal
of mass.  These issues motivate our use of controlled, idealized simulations to
achieve the required force ($\sim 10$ pc) and mass resolution ($\sim 1000 \,
\Msun$).

Recently, \citet{Penarrubia2012} highlighted the tension associated with
suppressing star formation in dwarfs while
simultaneously producing observable cores in their dark matter
distributions.  These authors primarily investigated the energy requirements for
creating constant-density cores in the density profiles of dwarf halos.
By contrast, we focus on the central masses of the dwarf spheroidals in this
work: we use idealized simulations to explore whether blow-out feedback of any
kind can realistically solve the TBTF problem in all the Milky Way dSphs,
including those with stellar masses as small as $\sim 10^5 M_\odot$.  We
predominantly examine the normalization problem pointed out by TBTF, rather than
the issue of the slope of the density profile that the cusp/core problem
implies.  

Our approach is to examine the effects of feedback on isolated dark matter halos
with peak circular velocities of $\sim 35$ km/s, the mass range associated with
TBTF halos \citep{Boylan-Kolch2012}.  We mimic baryonic feedback using an
externally tunable gravitational potential.  This allows us to test the amount
of gas that must be removed from the center of the halo in order to bring the
circular velocity into agreement with observations, as well as the energy
required to do so.  Our implementation also allows us to test whether cyclic
blowouts are effective at removing dark matter, as discussed by
\cite{Pontzen11}, and their relative efficiency compared to a single blowout of
the same total mass.

The layout of this work is as follows: in \S\ref{sec:sims}, we describe our
methods for producing the initial conditions and for emulating star formation
cycles, as well as present a resolution test; in \S\ref{sec:results}, we study
the dark matter distribution as a function of gas blown out and investigate the
energetic requirements; finally, in \S\ref{sec:discussion}, we discuss the
results, focusing specifically on the implications for the Too Big to Fail
problem.

\section{Simulations}
\label{sec:sims}
\subsection{Initial Conditions}
\label{sec:ICs}
Cosmological abundance matching models predict that galaxies with $L_V
\sim 10^5L_\odot$ form in dark matter halos with $V_\mathrm{max} \sim 35$ km/s
\citep{Quo2010}. Moreover, the five largest subhalos found in simulations of
Milky Way-size halos typically have $V_\mathrm{max} > 35$ km/s
\citep{Boylan-Kolch2012}.  This pinpoints halos with $V_\mathrm{max} \sim 35$
km/s as a characteristic size of concern.  Such a halo (with the circular
velocity curve peaking at a radius of 2.2 kpc, as expected for subhalos) is
shown in Figure~\ref{fig:ICs}; the points are circular velocity curve determinations
at the half light radii $r_{1/2}$ of each of the nine brightest Milky Way dSphs
\citep[taken from][who relied on data from the literature]{Wolf2010}.  Six of these nine have luminosities $L_V < 10^6$ L$_\odot$.
The data points are sized in proportion to their luminosities, which range from
$L_V=2.2 \times 10^5$ L$_\odot$ (Draco) to $1.7 \times 10^7$ L$_\odot$ (Fornax).
Their associated densities are clearly low compared to both the naive
expectations of abundance matching and the expected densities of the most
massive Milky Way subhalos.

With this as motivation, we initialize a dark matter halo with $V_\mathrm{max} =
35$ km/s at 2.2 kpc using a Hernquist \citeyear{Hernquist1990} sphere, which
follows the roughly-expected $\rho \propto r^{-1}$ dependence at small radius.
We do this by self-consistently sampling the phase space distribution function
of the model \citep[see also][]{kazantzidis04,Bullock2005,zemp08}. As long as
the resolution is appropriate, generating initial conditions in this manner can
produce systems that stay in equilibrium for over a Hubble time
\citep{kazantzidis04}. We have developed a code that applies this technique to
generate initial conditions for a variety of density profiles assuming isotropic
velocity dispersions. Our code, named \emph{spherIC}, is publicly
available\footnotemark and can also generate systems with an embedded stellar
component that follows its own density distribution, chosen from a set of
profiles typical for stellar systems.

\footnotetext{https://bitbucket.org/migroch/spheric}

\begin{figure*}
 \centering
   \includegraphics[width=0.49\textwidth]{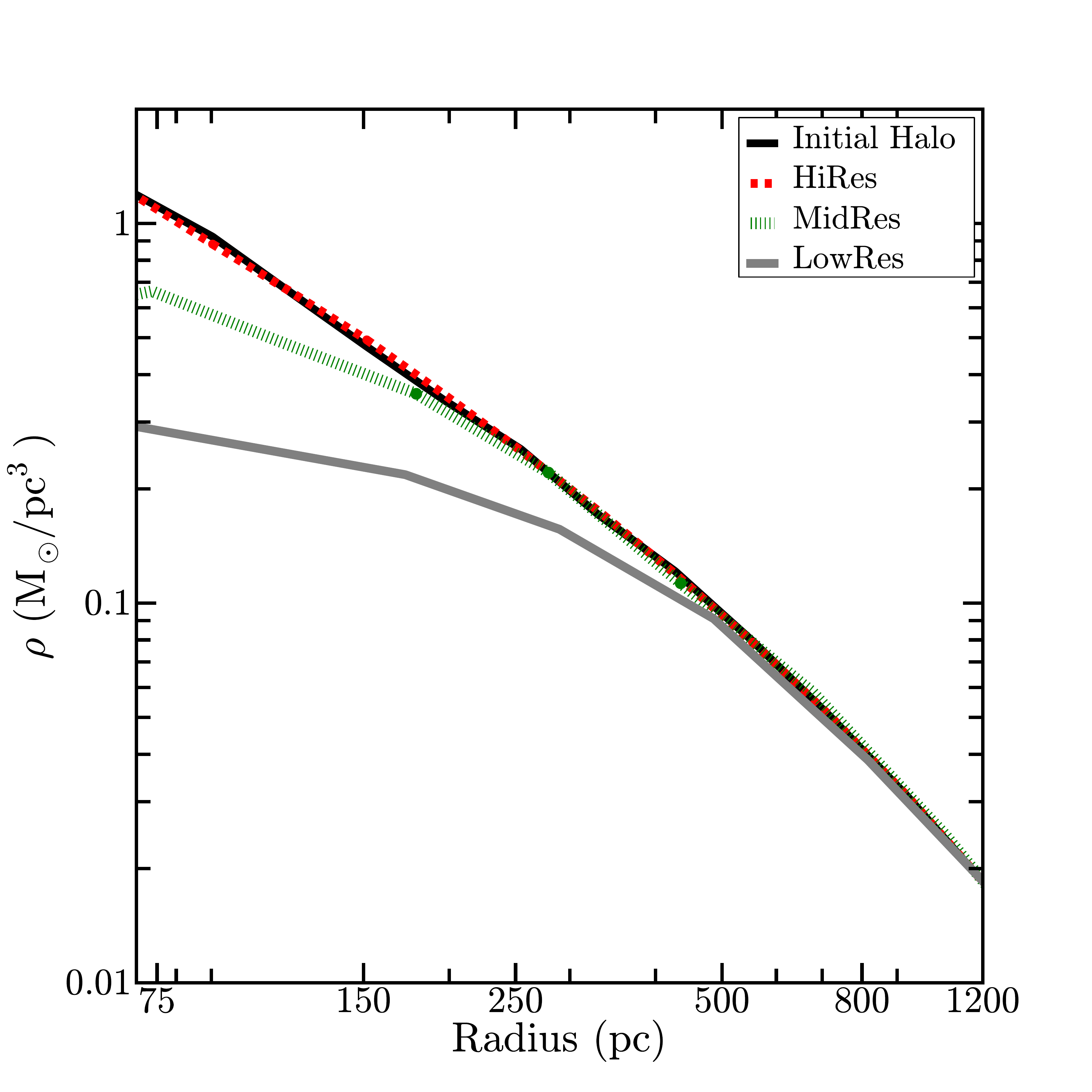}
 \centering
   \includegraphics[width=0.49\textwidth]{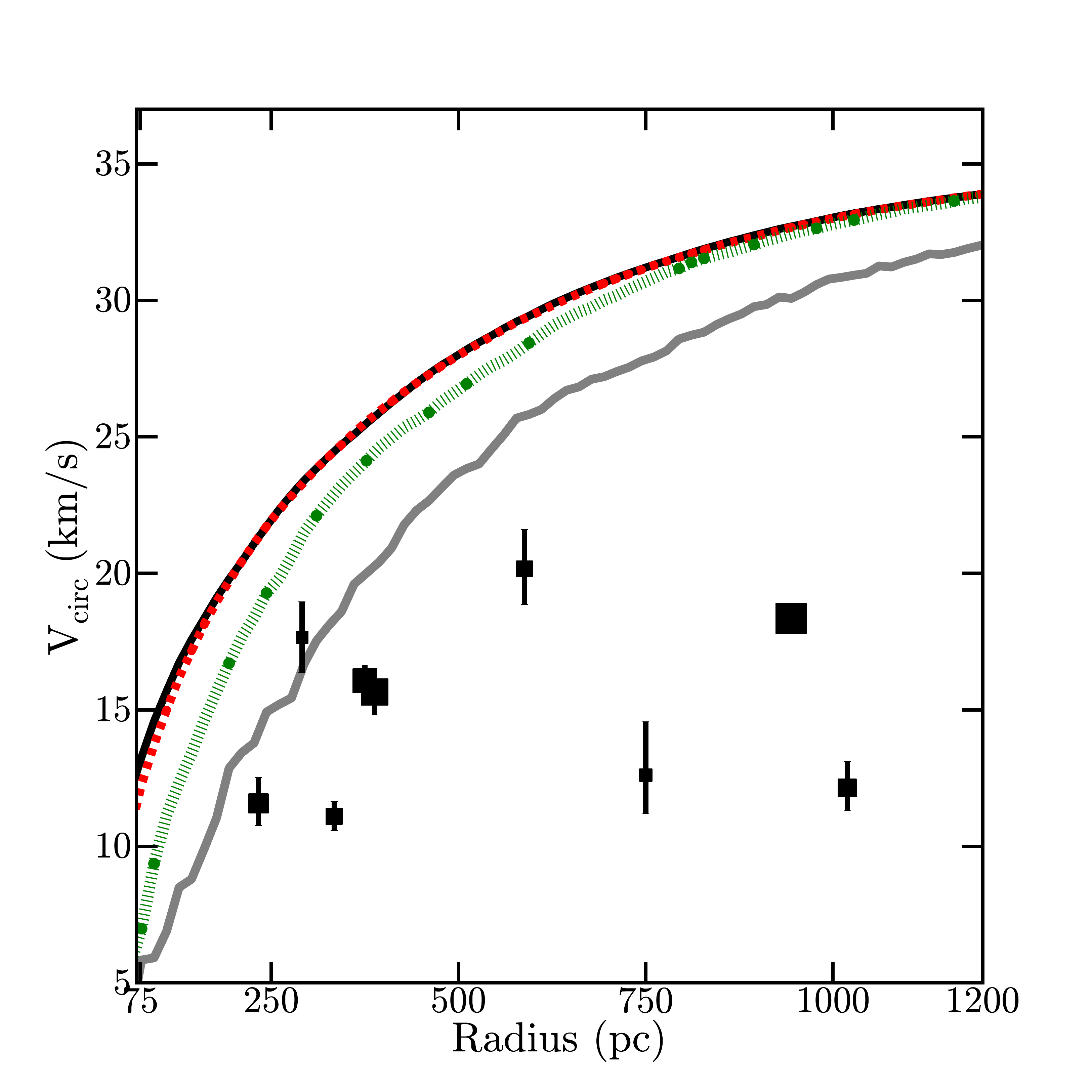}
   \caption{Resolution test.  Plotted are the density (left) and circular
     velocity (right) profiles for the isolated halo initially (solid black) and
     after running with no external potential for 5 Gyr with three different
     mass and force resolutions as labeled in the caption and in Table 1.  The
     highest resolution Milky Way cosmological simulations that have been run
     and can test feedback effects on dSph satellites have mass and force
     resolutions comparable to our LowRes runs
     though non-cosmological simulations are able to
     exceed the resolution of the MidRes run \citep{Teyssier2013}.  Although the
     density converges in the LowRes run at $\sim500$ pc, the circular velocity
     does not converge until beyond 1 kpc; to compare with the half-light radii
     of the dSph satellites, one requires convergence within $\sim 250$ pc, the
     smallest half-light radius of the Milky Way dwarfs.  Since only the HiRes
     run does not suffer from numerical effects in this region, we use this
     resolution exclusively for the experiments presented in the rest of this
     paper.}
\label{fig:restest}
\end{figure*}

In order to ensure that our results are stable to numerical effects, we simulate
our initial halo in isolation at increasing force ($\epsilon =$ 10, 70, 120 pc)
and mass ($m_p =$ 760, 24000, 150000 $\Msun$) resolution as detailed in
Table~\ref{tab:res}.  Figure~\ref{fig:restest} shows the resultant density and
circular velocity profiles after 5 Gyr for each of these runs.  The
highest resolution hydrodynamic simulations to date that study the formation and
evolution of a Milky Way-like halo and its dwarf satellites
have been run with force softenings comparable to our lowest resolution test
\citep[e.g.][]{Brooks2012,Zolotov2012}.  We see that the circular velocity curves for
runs at this resolution are under-resolved at all relevant radii. Though
numerical effects set in at $\sim4\epsilon$ in density, the cumulative circular
velocity remains divergent to larger radii.  \citet{Zolotov2012}, who examined
the TBTF problem in hydrodynamic simulations, limit their analysis to scales of
1 kpc or larger, where their mass profiles (circular velocity profiles) are
converged to 80\% (90\%). Smaller scales, $r < 1\,{\rm kpc}$, are most relevant for
the TBTF problem, however: all of the Milky Way dSphs have
$r_{1/2} \la 1$ kpc, with five \textless 500 pc and the smallest, Leo II, has
$r_{1/2} \sim 250$ pc.  To ensure that the circular velocity has converged at
these radii, the remainder of our work relies on simulations with 760 $\Msun$
particles and 10 pc force resolution, equivalent to the HiRes runs shown in
Figure~\ref{fig:restest}.

\begin{table}
\centering
\begin{tabular}{|l|c|c|c|}
\hline \textbf{Simulation}  & \textbf{$m_p$ ($M_\odot$)}  & \textbf{$\epsilon$ (pc)} & \textbf{$N_p$} \\ 
\hline\hline HiRes & $7.6\times10^2$ & 10 & 3,000,000 \\ 
\hline MidRes & $2.4\times10^4$  & 70 & 96,891 \\ 
\hline LowRes & $1.5\times10^5$ & 120 & 30,000 \\ 
\hline 
\end{tabular} 
\caption{Parameters of the runs used in the resolution test
  (Figure~\ref{fig:restest}) where $\epsilon$ is the Plummer-equivalent
  softening length.  The remainder of the simulations we discuss in this paper
  use the HiRes parameters.}
\label{tab:res}
\end{table}

\subsection{Modeling Gas Blowouts}
\label{sec:code}
We model a star formation cycle (i.e. gas accretion onto a central galaxy and
the subsequent ejection of that gas) by varying the properties of a
spherically-symmetric gravitational potential placed at the center of the halo.
Specifically, we added an externally tunable Hernquist potential to the N-body
code Gadget2 \citep{Springel2005} such that each particle has an additional
acceleration given by
\begin{equation}
 \vec{a} = \frac{-GM_\mathrm{gal}(t)}{[r+b(t)]^2}\frac{\vec{r}}{r},
\label{eqn:Hernaccel}
\end{equation}
where $M_\mathrm{gal}(t)$ is the total mass in the potential at time $t$ and
$b(t)$ is related to the half-mass radius of the potential, $r_{1/2}$, by
$r_{1/2} = b/(\sqrt{2}-1)$.  To prevent the acceleration of a particle from
becoming unphysically large when $r\to0$, we ``soften'' the potential by setting
$\vec{r}/r\to\vec{r}/\epsilon$ when $r<\epsilon$, the Plummer-equivalent
softening length. 

\begin{figure}
 \centering
  \includegraphics[width=0.45\textwidth]{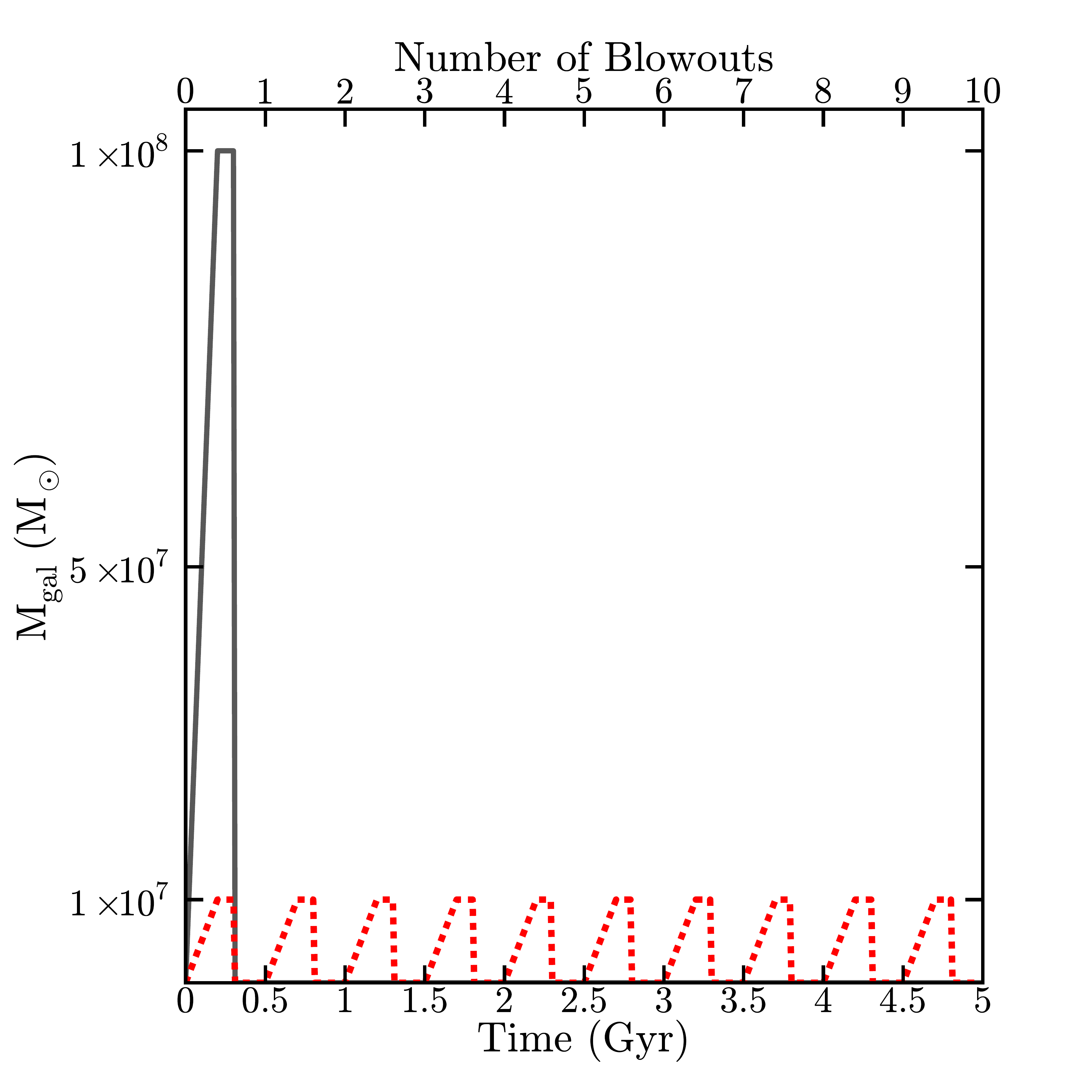}
  \caption{A representative example of our blowout scheme.  Plotted is the mass
    in the central potential as a function of time for two of our runs that blow
    out the same total amount of gas.  In grey is the mass as a function of time
    for a single blowout with $M_\mathrm{max}=10^8M_\odot$; the red dotted line
    shows the same for repeated blowouts with $M_\mathrm{max}=10^7M_\odot$.  A
    single cycle takes 500 Myr.  These two cases result in the same cumulative
    total of mass displaced, but as shown in Figure~\ref{fig:threemasses}, the
    single large burst affects the dark matter density to a larger extent. }
\label{fig:galmass}
\end{figure}

Our implementation allows us to specify the properties of the potential
($M_\mathrm{gal}$ and $r_{1/2}$) at any time.  Our fiducial runs fix $r_{1/2}$
and vary $M_\mathrm{gal}$.  Most of our runs force $r_{1/2} = 500$ pc, which is a
typical half-light radius among the bright Milky Way dSphs.  We vary
$M_\mathrm{gal}$ over a series of cycles, with a fiducial period of $500$ Myr
(see Figure~\ref{fig:galmass}).  Specifically, $M_{\rm gal}$ grows linearly from
zero over 200 Myr to its maximum mass, $M_\mathrm{max}$, then remains constant
for 100 Myr to allow the halo to come to equilibrium.  We then mimic a blowout
by forcing $M_\mathrm{gal}$ to instantaneously return to zero, where it remains
for 200 Myr before beginning the next cycle.  We have tested a number of other
models for blowouts, including those with different periods, models without a
relaxation time, sinusoidal modulations, and models with $M_\mathrm{gal}$ constant
and $r_{1/2}$ varying.  The model we show here produces the maximal effect on
the rotation curve, though the qualitative results are very similar in most
cases.  The only exception is the sinusoidal model, which is symmetric and
effectively produces no change in the density distribution.   We have also 
tested a cylindrically symmetric potential, and found qualitatively similar 
results to the spherical cases.

In what follows we present results for models with $M_\mathrm{max} =
10^6M_\odot$, $10^7M_\odot$, $10^8M_\odot$, and $10^9M_\odot$.  For each of
these we vary $M_\mathrm{gal}$ from zero to $M_\mathrm{max}$ and back to zero
ten times over a total of $5$ Gyr.  We output snapshots after every blowout, so
that we can investigate the effect of any number of star formation cycles on the
associated dark matter profile.  For example, the grey line in
Figure~\ref{fig:galmass} illustrates the galaxy mass as a function of time over
one cycle of $10^8M_\odot$ while the red line shows ten cycles of $10^7M_\odot$
each -- in both of these runs, a total of $10^8M_\odot$ is blown out from the
halo.  We also test how strongly the results depend on the scale radius by
presenting new runs with $r_{1/2} = 100$ pc and $M_\mathrm{max}=10^7M_\odot$
and $10^8M_\odot$.


\section{Results}
\label{sec:results}
\begin{figure*}
 \centering
    \includegraphics[width=0.49\textwidth]{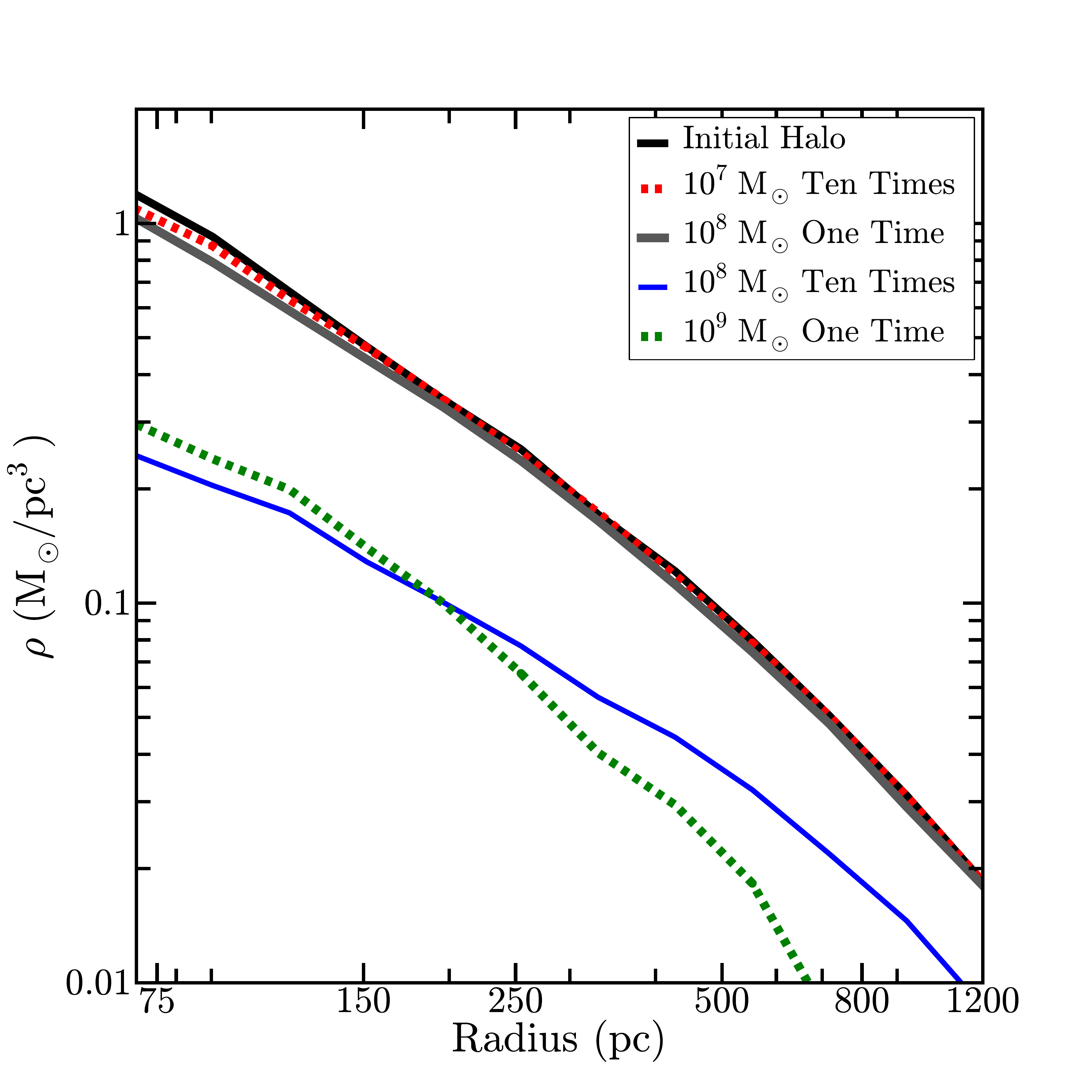}
 \centering
    \includegraphics[width=0.49\textwidth]{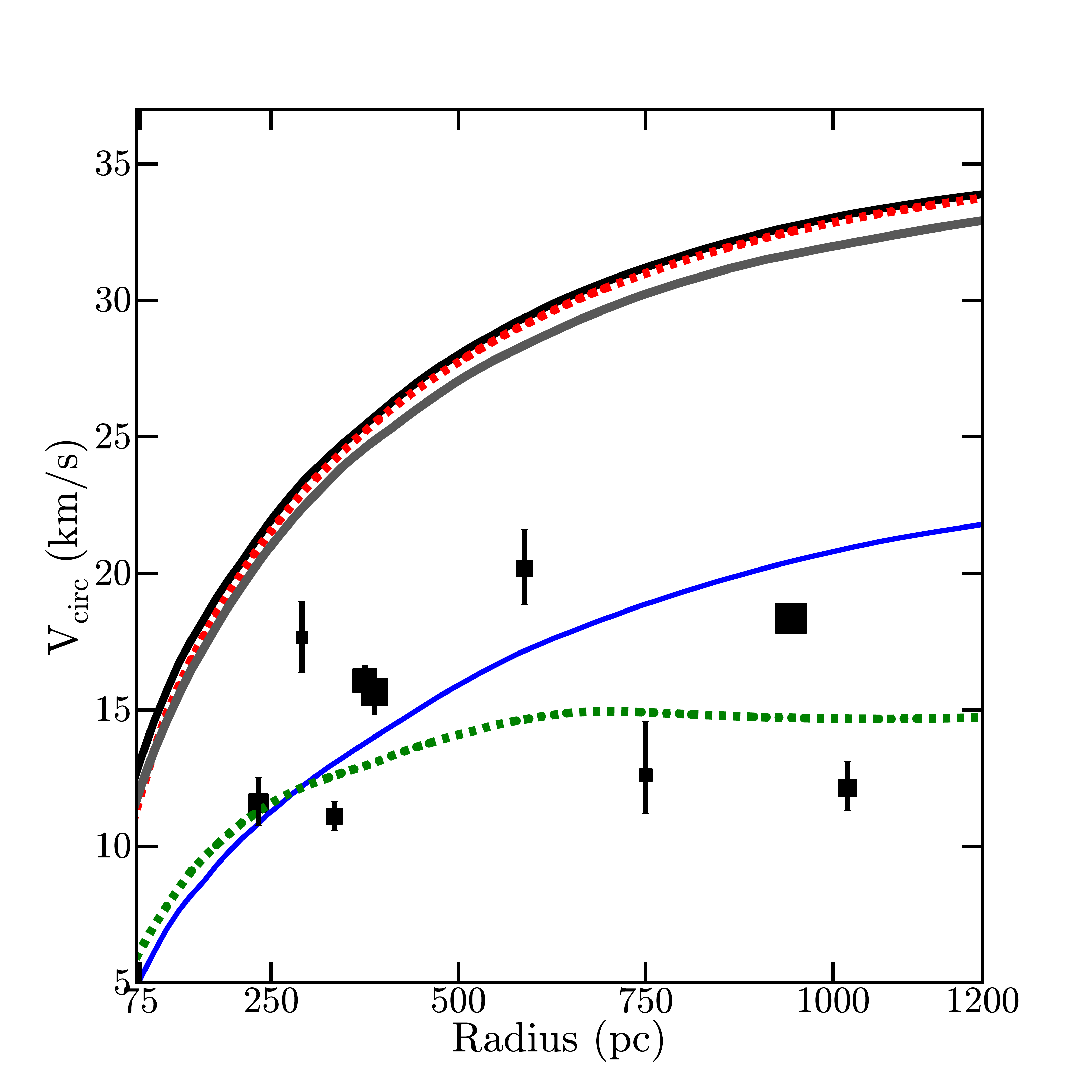}
    \caption{The density (left) and circular velocity (right) profiles of the
      halo after ten blowouts of $10^7M_\odot$ and one blowout of $10^8M_\odot$
      (upper lines; as illustrated in Figure~\ref{fig:galmass}), and after ten
      blowouts of $10^8M_\odot$ and one blowout of $10^9M_\odot$ (lower lines),
      all with $r_{1/2} = 500$ pc (the qualitative results are similar for
      $r_{1/2} = 100$ pc).  Removing $10^7M_\odot$ ($10^8M_\odot$) ten times is
      less effective at removing dark matter from the inner region of the halo
      than removing $10^8M_\odot$ ($10^9M_\odot$) once.  Furthermore, note that
      $\sim10^9M_\odot$ of baryons must be removed from the galaxy -- though it
      does not have to leave the halo -- to bring the circular velocity into
      agreement with the data.}
\label{fig:threemasses}
\end{figure*}

Figure~\ref{fig:threemasses} shows changes in the density and circular velocity
profiles of our initial halo after undergoing blowout(s) of various masses.  We
directly compare ten blowouts of $10^7M_\odot$ ($10^8M_\odot$) to one blowout of
$10^8M_\odot$ ($10^9M_\odot$), and find that for both values of
$M_\mathrm{max}$, a single blowout (grey line in Figure~\ref{fig:galmass})
removes more dark matter from the center of the halo than repeated blowouts that amount
to the same total baryonic mass cycled through the halo (red line in 
Figure~\ref{fig:galmass}).  While we do see some evidence that cyclic, lower mass 
blowouts remove mass preferentially from the inner regions compared to a more massive 
blowout, being more effective at forming a ``core," the density distribution never
becomes perfectly flat in the center -- some degree of cuspiness always remains.

Figure \ref{fig:threemasses} illustrates that in order to bring our fiducial
TBTF halo (solid black) into agreement with the density of a typical dSph, a
total of $\sim10^9M_\odot$ of material must be cumulatively ejected from the 
halo in either one massive blowout (dashed green, the biggest effect) or in 
a few repeated smaller blowouts (solid blue) totaling this amount.  
This mass exceeds the entire baryonic allotment for a field halo of
$M_{\rm vir} \simeq 5 \times 10^{9} \Msun$, which is the mass associated with an
$M_\star \simeq 10^6 \Msun$ galaxy according to the extrapolated abundance
matching of \citet{Behroozi12} and also the virial mass associated with a
typical TBTF halo at the time of infall \citep{Boylan-Kolch2012}.  This suggests
that if feedback is responsible for the change in the density profile, it must
be cyclic, such that a baryonic mass element may be reused in repeated blowouts.
 
\begin{figure*}
 \centering
    \includegraphics[width=0.49\textwidth]{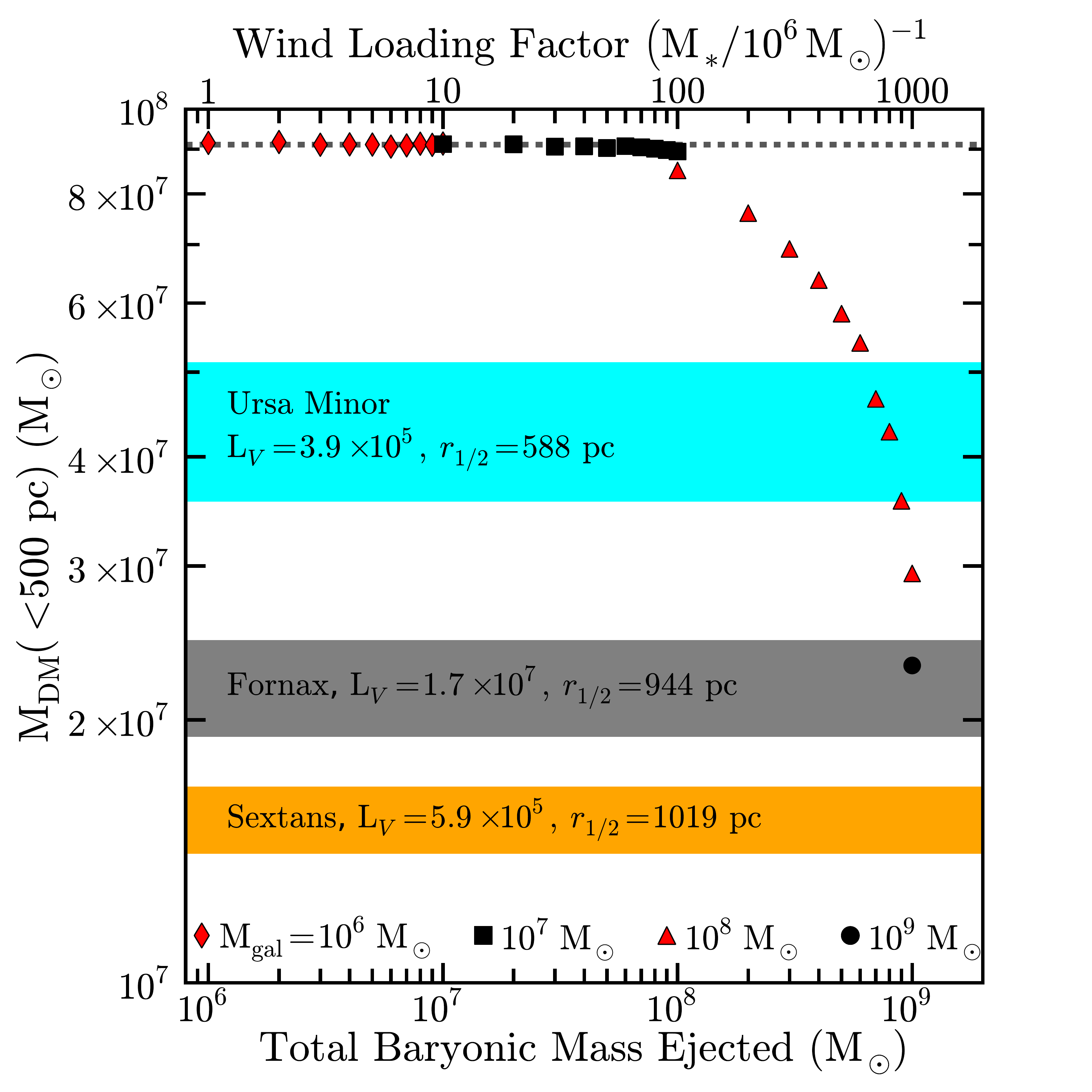}
  \centering
	\includegraphics[width=0.49\textwidth]{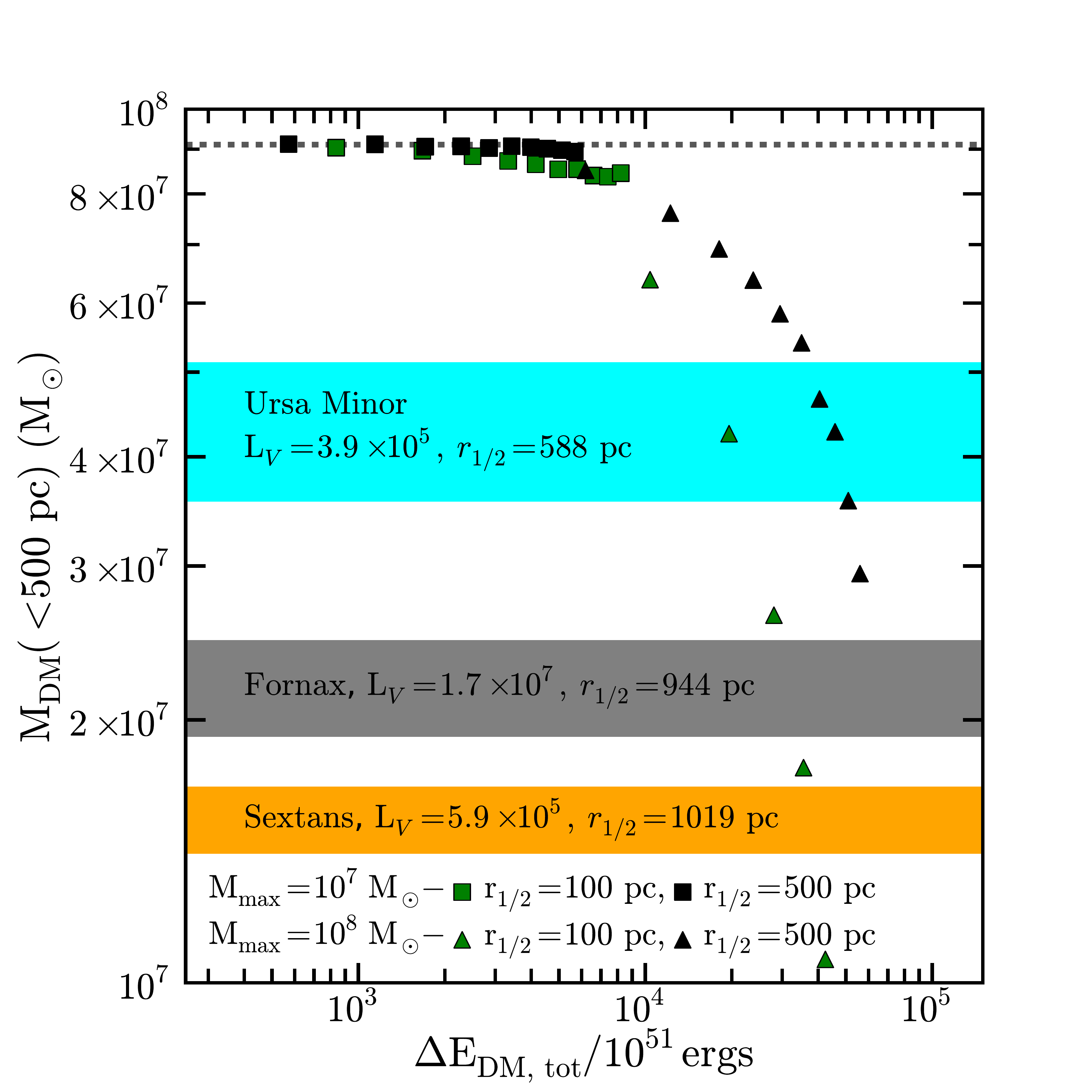}
    \caption{\textit{Left}: The mass remaining within 500 pc after repeated
      blowouts of a galaxy with $M_\mathrm{max} = 10^6$, $10^7$, $10^8$, and
      $10^9M_\odot$, all with $r_{1/2}$ = 500 pc, as a function of mass blown
      out. \textit{Right}: The mass remaining after blowouts of $10^7M_\odot$
      and $10^8M_\odot$ with either $r_{1/2} = 100$ pc (in green) or 500 pc
      (in black) as a function of the cumulative change in the dark
      matter energy.  The dotted line indicates the original mass within 500 pc,
      and the colored bands indicate the dark matter within 500 pc for Ursa Minor, 
      Fornax, and Sextans.  As
      the stellar component of Fornax contributes non-negligibly to the mass
      near its center, we have subtracted $7\times10^6M_\odot$ from the
      dynamical mass in order to account for the stellar mass within $500$ pc
      for this galaxy \citep{Jardel2012}.  More than several times
      $10^8M_\odot$ must be ejected to bring the mass into agreement with Ursa
      Minor (though each blowout may be $\sim 10^8 M_\odot$), and the requisite 
      energy also exceeds the total supernovae budget for
      six of the nine classical dSphs.  Furthermore, we note that $\Delta
      \mathrm{E_{DM}}$ is a lower limit on the energy that must be injected, as
      it does not account for energy escaping via radiation or the energy
      required to eject the baryons.}
\label{fig:sumresults}
\end{figure*}

A more general presentation of our results is given in
Figure~\ref{fig:sumresults}.  Each panel shows the dark matter remaining within
500 pc after multiple blowout runs.  On the left, we present the mass of dark
matter remaining as a function of total mass ejected and on the right we show
the same quantity as a function of the total energy added to the dark matter
(see below).  For reference, the horizontal dotted line shows the initial dark
matter mass within 500 pc and the shaded horizontal bands show estimates of the
dark matter mass within 500 pc for three representative dwarfs, as determined in
\citet{Boylan-Kolch2012}.

In the left panel of Figure \ref{fig:sumresults}, the different symbol types
correspond to different values of the mass blown out per cycle, spanning
$10^6M_\odot$ to $10^9M_\odot$ as indicated in the figure.  Multiple points with
the same symbol type correspond to repeated blowouts of the same mass.  The
points here are from runs with $r_{1/2} = 500$ pc.  As discussed above, a single
massive blowout removes more mass from the center of the halo than the
cumulative effect of 10 smaller blowouts that result in the same total mass
expelled.  For reference, the upper axis shows the implied mass loading factor,
normalized for a dSph with $M_* = 10^6 \, M_\odot$.  We see that a minimum of $7
\times 10^8 \Msun$ of material must be removed in order to reach the observed
density of the densest dwarfs shown, Ursa Minor (cyan band), though each
individual blowout cycle may be far less massive. This would imply a mass
loading factor of $\sim 1000$ if we assume a stellar mass-to-light ratio of
$M_\star/L_V=2$ for this system.  We note that we are defining the mass loading factor as the mass removed
from the {\em galaxy} divided by the mass formed in stars.  This number is the same whether or not
a gas parcel is lost from the halo entirely or if it eventually falls back into the galaxy and is blown out multiple times.

The mass of Fornax is represented by the grey band.  Though the density of
Fornax is significantly lower than that of Ursa Minor, it may be the easiest
dwarf to explain because its reservoir of stars is much greater ($M_\star \simeq
4 \times 10^7$).  We see that a cumulative expulsion of $\sim 10^9 \Msun$ can in
principle match the central density of Fornax, which would require a more modest
-- though still large -- cumulative mass-loading of $\sim 25$.  However, this is 
only one system and it does not explain the unexpectedly low densities of the 
other, less luminous dwarfs.

Another way to characterize the problem inherent in lowering the densities in
the faintest dwarfs is to consider the energy required to bring densities into
accordance with observations \citep{Penarrubia2012}.  The right hand panel of
Figure~\ref{fig:sumresults} presents the mass remaining within 500 pc as a
function of the cumulative energy injected into the dark matter after a series
of $10^7M_\odot$ (squares) and $10^8M_\odot$ (triangles) blowouts for two values
of $r_{1/2}$.  Green symbols correspond to $r_{1/2} = 100$ pc blowouts and the
black symbols correspond to $r_{1/2} = 500$ pc blowouts.  The smaller
$r_{1/2}$ runs produce marginally bigger effects for the same energy.  However,
the three dwarfs shown by bands in Figure~\ref{fig:sumresults} have $r_{1/2}
\simeq 600$ pc, $900$ pc, and $1000$ pc, respectively.  Thus we regard our $100$
pc blowout models as quite conservative limits.

We calculate the energy injected into the dark matter by measuring the energy
difference in the dark matter before and after each blowout.  We ignore the
energy ``lost" when the dark matter re-contracts in response to central
potential regrowth.  This is because we are interested in the energy imparted to
the dark matter by explosive feedback, which has nothing to do with how the gas
falls back in to regrow the central galaxy (ignoring this component amounts to
changes in the presented values at the factor of $\sim 2$ level).  For our
fiducial runs with $r_{1/2} = 500$ pc, we see that more than $4 \times 10^{55}$
ergs of energy must be delivered to the dark matter before the inner mass becomes
consistent with Ursa Minor.  Assuming an energy per explosion of $E_{SN} =
10^{51}$ erg, this corresponds to more than $40$,$000$ supernovae worth of
energy injected directly into the dark matter with 100\% coupling.  Given that
we expect approximately one SNII explosion per 100 $M_\star$ formed for a
typical IMF \citep{Kroupa2002}, this exceeds the total available energy budget
for all of the type II supernova that have occurred in Ursa Minor.  Indeed, it
exceeds the associated supernovae budget for six of the nine galaxies of concern
in Figure 1, all of which, according to extrapolated abundance matching, should
be sitting in massive halos.  The three most luminous dSphs may be in the range
of viability, but we must assume that the energy couples directly to the dark
matter, ignoring the energy required to expel the gas from the halo and
radiative losses.  The real energetic requirements may be more than a factor of
10 larger \citep{Creasey2012}.  We also note that a single, large blowout
injects more energy into the dark matter than is imparted by ten successive, smaller
blowouts of the same total mass.

As discussed above, Fornax (with $r_{1/2} \simeq 1$ kpc and $L_V \simeq 1.7
\times 10^7 L_\odot$) appears to be the best candidate for having its density
lowered significantly by feedback effects.  We expect Fornax to have had $\sim 3
\times 10^5$ supernovae explosions over its history.  According to
Figure~\ref{fig:sumresults}, a $\sim 20\%$ coupling of $E_{SN}$ to the dark
matter could in principle have lowered the central density of a 35 km/s halo
enough to match the observed density of Fornax.  Interestingly, however, if we
run multiple blowouts at $\sim 10^9 \Msun$, we find that our host halo becomes
unbound all together.  This suggests that even in cases where the required
blowout is plausible energetically, there is something of a fine-tuning problem:
if feedback is really as effective as required, then many of these halos may
be destroyed all together.  If this level of coupling is generic, one might
expect slightly more star-rich galaxies to not exist at all.  Alternatively, 
the presence of these galaxies may indicate that star formation is strongly
suppressed by flatter central densities, such that the changing potential 
regulates further outflows; however, more detailed tests are necessary to
determine the strength of such a feedback loop, if it exists.

\begin{figure}
\centering
\includegraphics[width=0.45\textwidth]{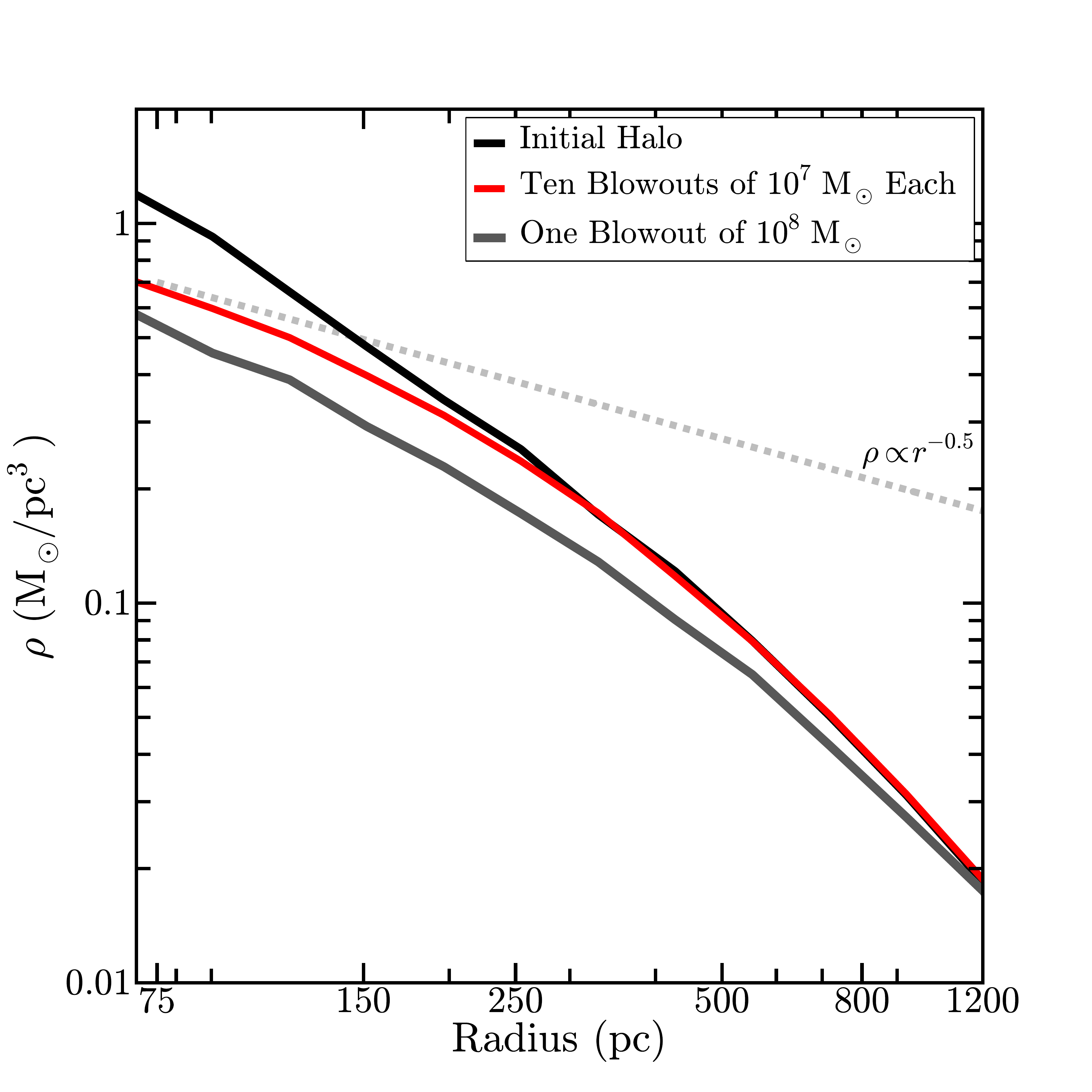}
\caption{The density profile of the halo after ten blowouts of $10^7M_\odot$ and
  one blowout of $10^8M_\odot$ with $r_{1/2}$ = 100 pc.  As with blowouts with
  $r_{1/2}$ = 500 pc, repeated blowouts are less effective than a
  single, more massive, blowout at displacing dark matter.  However, repeated
  small blowouts from the very center of the halo do begin to flatten the inner
  density profile, whereas a more massive potential removes mass even at $10
  r_{1/2}$.}
\label{fig:r100comp}
\end{figure}

One of the main results presented above is that repeated, cyclic blowouts do not
help in lowering the central densities of galaxies compared to single bursts.
However, we find that for a fixed amount of mass expelled from the galaxy (and
possibly cycled through the halo) cyclic blowouts preferentially remove dark
matter mass from the centers of halos, making them more effective at shallowing
cusps.  We find that the effect is most dramatic for the smallest $r_{1/2}$
runs.  Figure \ref{fig:r100comp} compares the density profile after ten blowouts
of $M_\mathrm{max} = 10^7M_\odot$ to one blowout with $M_\mathrm{max} =
10^8M_\odot$, both with $r_{1/2} = 100$ pc.  We see that several small blowouts
begins to flatten the density profile at the center of the halo, whereas one
large blowout displaces mass more evenly at all radii.  Though a thorough
investigation of this is outside the scope of this work, there does appear to be
evidence that the scheme proposed by \cite{Pontzen11} can lead to more core-like
dark matter profiles, perhaps consistent with those derived by
\citet{Walker2011}; however, it appears unlikely that it can affect the total
mass within the stellar extent at the level required to resolve the TBTF
problem. In practice, we are never able to produce true constant-density cores.
Rather we find mild cusps, $\rho \propto r^{-\alpha}$, with $\alpha \gtrsim 0.5$
-- significantly steeper than those observed in dSphs by \cite{Walker2011} and
\cite{Amorisco2012}.

\section{Conclusions}
\label{sec:discussion}
In this paper we have used a series of idealized numerical experiments to
investigate whether blowout feedback can plausibly resolve the Too Big to Fail
(TBTF) problem for subhalos seen in $\Lambda$CDM simulations
\citep{Boylan-Kolch11,Boylan-Kolch2012}.  We relied on a tunable central
potential to mimic the effects of cyclic baryonic feedback events within a
$V_{\rm max} = 35$ km/s halo -- the mass scale of concern for the TBTF
problem. 

Our overall conclusion is that supernovae feedback appears to be incapable of
solving the TBTF problem.  More specifically, our findings are as follows:

\begin{itemize}

\item In order to bring massive subhalo densities in line with those observed
  for Milky Way dSphs, a total of $\sim 10^9 \, \Msun$ of material must be
  ejected from the galaxy (though not necessarily from the halo).  This requires
  wind loading factors in excess of $\sim 500$ for the majority of
  Milky Way satellites.

\item Our fiducial feedback models that match the observed densities of Milky
  Way dwarfs require the deposition of $>$40,000 supernovae worth of energy
  directly into the dark matter with 100\% efficiency.  For typical initial mass
  functions, this exceeds the expected number of Type II supernova explosions
  for six of the nine brightest dSph satellites.  The most plausible exception
  is Fornax, with a density that may be explained with a $\sim 20 \%$ coupling
  of its full allotment of SN energy directly to the dark matter.  If this were
  the case, it might pose a fine-tuning problem for somewhat more luminous
  galaxies, as they might be expected to completely unbind their host halos.

\item Repeated blowouts are less effective at lowering the central densities of
  dark matter halos than a single blowout of the same cumulative mass and
  similar total energy imparted (see Figure 5). Repeated small bursts do produce
  shallower central cusps than a few large bursts.  However, we are unable to
  produce a true constant-density core from cyclic blowouts, even in the most
  extreme cases.  Importantly, the overall density remains higher at all radii
  when the same total mass is blown out in a few smaller events (see Figure 6).
  We conclude that, per unit energy delivered or per cumulative mass removed,
  cyclic blowouts are less effective than a single large blowout at reducing
  central densities of dark matter halos.

\end{itemize}

This work has focused on the effects of internal feedback on the density
structure of dark matter halos that are similar to those that will become
massive subhalos in Milky Way-mass halos at redshift zero.  We have not
considered the effects of subsequent evolution, including stripping from ram
pressure and tides, that may be important for some Milky Way subhalos
\citep{Read2006,Brooks2012,Zolotov2012,Arraki2012}.  Indeed, \citet{Arraki2012}
have shown that, while tidal evolution alone is insufficient to bring the
simulated subhalo population into agreement with observations of the MW dSphs,
it may be sufficient to produce the requisite changes in a subhalo that has
undergone adiabatic expansion due to baryon removal as a result of ram-pressure
stripping.  We note, however, that several Milky Way dSphs do not seem to have
had close pericentric passages with the Galaxy \citep{Lux2010,Sohn2012}, which
calls into question the influence of tides on their mass distributions.

In light of the uncertainties associated with environmental influences on dark
matter halo structure, the results presented here point to isolated, low-mass
galaxies as particularly important objects for testing the predictions of
CDM-based models.  Future optical and radio surveys will be capable of detecting
objects with stellar masses similar to the MW dSphs outside of the Local Group;
comparing their density structure to predictions from simulations will be
particularly enlightening.

\vskip1cm

\noindent {\bf{Acknowledgments}} \\
This work was funded in part by NSF grants AST-1009999, AST-1009973, and 
NASA grant NNX09AD09G.  M.B.-K. acknowledges support from the Southern California 
Center for Galaxy Evolution, a multi-campus research program funded by the University 
of California Office of Research.   J.S.B. was partially supported by the Miller Institute for 
Basic Research in Science during a Visiting Miller Professorship in the Department of 
Astronomy at the University of California Berkeley.  The authors thank Manoj 
Kaplinghat for insightful discussions, and also thank Alyson Brooks, Arianna di Cintio,
Avishai Dekel, Anatoly Klypin, Andrea Maccio, Jorge Pe\~narrubia, Andrew Pontzen,
Antonio Del Popolo, Chris Purcell, Justin Read, Romain Teyssier, Erik Tollerud, 
Matthew Walker, and particularly the anonymous referee for many helpful comments.

\bibliographystyle{apj}
\bibliography{blowout_paper}
\label{lastpage}
\end{document}